\documentclass{article}
\usepackage{aip}

\input{tcilatex}

\begin{document}

\title{The connection between non-exponential relaxation and fragility in
supercooled liquids }
\author{V. Halpern \\
Department of Physics, Bar-Ilan University, Ramat-Gan 52900, Israel.\\
e-mail: halpeh@mail.biu.ac.il}
\maketitle

\begin{abstract}
Among the outstanding problems in the theory of supercooled liquids are the
reasons for the rapid increase in their viscosity and relaxation times as
the temperature is lowered towards the glass transition temperature $T_{g}$,
the non-exponential time dependence of the relaxation, and the possible
connection between these two properties. The ferromagnetic Potts model on a
square latice is a simple system that is found to exhibit these properties.
Our calculations show that in this system the connection between them is
associated with the dependence on temperature and time of the average
environment of the sites. Some of the consequences of this for understanding
the behavior of supercooled liquids are discussed.
\end{abstract}

\section{Introduction}

There are a number of features common to most supercooled glass-forming
liquids near their glass transition temperature $T_{g}$, and which differ
appreciably from those of normal liquids and of crystalline solids \cite%
{Angell-rev}. Two of the most prominent of these are the temperature
dependence of the properties close to $T_{g}$ and their time dependence. In
many supercooled liquids it is found that as the temperature is lowered
towards $T_{g}$ properties such as the viscosity $\eta $ and the mean
dielectric and mechanical relaxation times $\tau $ increase much more
rapidly than the Arrhenius behavior associated with a fixed activation
energy, which would give $\ln (\tau /\tau _{0})=E/(kT)$. According to
Angell's classification scheme \cite{Angell-fragile}, strong liquids are
those in which the departure from the Arrhenius behavior is slight and
fragile ones are those for which the temperature dependence departs strongly
from the Arrhenius behavior as the temperature is lowered towards $T_{g}$.
The time dependence of the main ($\alpha $) relaxation process in
supercooled liquids is also not described by a simple exponential function,
but can often be well approximated in the time domain by a stretched
exponential function, 
\begin{equation}
\Phi (t)=\exp [-(t/\tau )^{\beta }],  \tag{1}
\end{equation}%
where $\beta $ is called the stretching exponent, $0<\beta \leq 1$, and the
degree of non-exponentiality is given by $n=1-\beta $. These two properties
seem to be connected, since $n$ tends to be much smaller in strong glasses
than in fragile ones \cite{Ngai-rev}. While such a connection is found in
various model systems, these do not usually provide any simple explanation
of it in terms of the microscopic properties of the system.

In this paper, we discuss a simple model system for which we find these
properties, and also a simple explanation of them and their connection in
terms of the average environments of the sites, namely the ferromagnetic $q$%
-spin Potts model \cite{Wu-rev}\ of spins that can take any of $q$ distinct
values located on the sites of a lattice. At first sight, there seems to be
very little resemblance between such a model of spins on a lattice and
supercooled liquids (unless these spins are assumed to represent regions of
the system with different properties, as in the facilitated kinetic Ising
model of Fredrickdson and Andersen \cite{FA}, for instance). However, it is
well known that the dielectric relaxation of plastic crystals such as
ethanol is very similar to that of the corresponding supercooled liquid \cite%
{Benkhof}, so that spatial disorder is not an essential requirement for the
typical behavior of supercooled liquids. For these plastic crystals, one
widely-used model is that of thin rods rotating about their centres which
are fixed on a lattice \cite{Bermejo}. An alternative model for these
materials would be of rods rotating between a number of discrete
orientations, which could be represented by spins as in the Potts model. In
that case, the Potts model corresponds to a different (and in principle no
less plausible) interaction between adjacent molecules, and so its
properties may also be expected to resemble those of plastic crystals and
supercooled liquids.

While the ferromagnetic Potts model has been extensively studied for over 50
years, most studies have been associated with its thermodynamic rather than
its relaxation properties. Even a recent study of the latter was concerned
(like most recent papers) with a distribution of positive and negative
values of the interaction $J$ between equal spins on adjacent sites \cite%
{Brang}. The system that we study is the much simpler ordered ferromagnetic
Potts model, in which the interaction $J$ has the same positive value
between any pair of equal spins on adjacent sites. A recent study of the
dynamic properties of this ferromagnetic model \cite{Velytsky} was concerned
with the dynamics of the phase transition rather than with relaxation some
distance from it which we study. Our motivation for studying the Potts model
were based on the following two observations about non-Arrhenius behavior
and non-exponential relaxation. Firstly, both simple exponential relaxation
and a fixed activation energy are associated with a fixed environment for
the molecules, so that departures from both these types of behavior will
occur if the environment of the molecules changes. Secondly, just as
stretched exponential relaxation means that as time proceeds the relaxation
rate decreases, so does the increase in the effective activation energy as
the temperature is lowered mean that the relaxation rate is less than it
would be for a fixed environment. In the ferromagnetic Potts model, as the
temperature is lowered towards the phase transition temperature $T_{c}$,
even before the critical region close to $T_{c}$ is reached, larger and
larger clusters of spins of the same size are expected to be formed, so that
the environment of the spins changes to make transitions more difficult, as
discussed in section 4. Similarly, at any temperature those spins that are
inside a cluster of identical spins will have a lower probability of making
a transition, so that as time proceeds increasing fractions of the spins
that have not made a transition will be in such clusters, and this leads to
a slowing down of the transition rate. Thus, the Potts model should provide
a simple system in which, by tracking the changes in the environments of the
sites, we can see a common source for non-Arrhenius behavior and
non-exponential relaxation.

The idea of the importance of a molecule's local environment is not new, and
has been considered previously by quite a number of authors. For instance,
Ngai's coupling model essentially involves changes in the coupling of a
molecule to its surroundings as time proceeds \cite{Ngai-coupling}. The
defect diffusion model of Bendler and his co-workers \cite{Bendler} can also
be regarded as involving changes in the environment of free defects
surrounding a molecule. In the theory of the fragility of supercooled
liquids proposed by Tanaka \cite{Tanaka}, the $\alpha $-relaxation is
associated with the dynamics of the formation and annihilation of solid-like
islands, but he considers only its temperature dependence and not the time
dependence of the relaxation. Such islands are examples of local
inhomogeneities or heterogeneities, and these have been considered by
Garrahan and his co-workers \cite{Garrahan} in systems with dynamic
constraints. Numerous other approaches have been based on 2-state or
2-region models and dynamic constraints, such as the much studied
Fredrickson-Andersen model \cite{FA}. However, none of these approaches
examined model systems with no externally imposed assumptions, such as the
Potts model considered in this paper.

A detailed description of the $q$-spin ferromagnetic Potts model and its
relaxation properties is presented in section 2, while section 3 contains
the results of extensive computer simulations on it. These results are
discussed and explained in section 4, and their implications for real
supercooled liquids are then considered, while our conclusions are presented
in section 5.

\section{The ferromagnetic $q$-spin Potts model.}

The Hamiltonian for the ordered ferromagnetic $q$-spin Potts model with
interactions only between the spins at adjacent sites can be written as \cite%
{Wu-rev} 
\begin{equation}
H=-J\sum_{i}\sum_{j(i)}\delta (\sigma _{i},\sigma _{j})  \tag{2}
\end{equation}%
where $J>0,$ the first sum is over all the sites $i$ in the system and the
second one over all the sites $j(i)$\ that are nearest neighbors of the site 
$i$, the spins $\sigma _{i}$ can take any integer value between $1$ and $q$,
and $\delta $ is the Kronecker delta. Hence, the energy associated with a
spin having $z$ adjacent sites with the same spin is just $-zJ$. The
probability of a change in the spin at a site which involves an increase of
energy $\Delta E$ at temperature $T$ was taken to have the standard form 
\begin{eqnarray}
w &=&w_{0},\quad \Delta E<0  \TCItag{3} \\
&&w_{0}\exp (-\Delta E/k_{B}T),\quad \Delta E>0  \nonumber
\end{eqnarray}%
We note that the Potts model incorporates temperature explicitly, unlike
models with dynamic constraints such as that of a system of rotating rods
which has been applied to materials such as ethanol and in which the
temperature had to be introduced in terms of the lengths of the rods \cite%
{Bermejo}.

A question that always arises in simulations is whether to update the spins
one at a time, which may be more realistic but requires a lot of computer
time, or to update all of them simultaneously in one time step, which
usually requires much less computer time but is not necessarily very
realistic. For our model, which involves only interactions between spins on
adjacent sites, we use the fact that the sites of the square lattice can be
chosen as $\mathbf{n}=$ $n_{1}\mathbf{a}_{1}+n_{2}\mathbf{a}_{2}$, where $%
\mathbf{a}_{1}=(1,0)$ and $\mathbf{a}_{2}=(0,1)$. This lattice can be
divided into four sub-lattices, with $%
(n_{1},n_{2})=(2j,2k),(2j+1,2k),(2j,2k+1),$ and $(2j+1,2k+1)$ where $j$ and $%
k$ are integers, such that adjacent sites are not on the same sub-lattice.
Accordingly, the simultaneous updating of the spins on one sub-lattice at a
time corresponds physically to updating the spins one at a time, but with a
saving in computer time comparable to that obtained by updating all the
spins at once. At each time step all the spins on one sub-lattice were
updated simultaneously, while the sub-lattices were considered in turn.
Because of possible problems with the Metropolis algorithm, in which the
final state at each step is chosen at random, we used the equivalent of the
continuous time Monte-Carlo algorithm \cite{Gotcheva}, and considered at
each step all possible transitions from each site with their appropriate
probabilities.

The ferromagnetic $q$-spin Potts model on a square lattice for an infinite
system exhibits a phase transition at a critical temperature $T_{c}$, and
this transition is of second order for $q\leq 4$ and of first order for $q>4$%
. In terms of the variable $v=\exp (J/[k_{B}T])-1$, the values of $T_{c}$
are given by the solutions $v=v_{c}$ of the equation $v^{2}=q$ \cite{Baxter}%
. For our simulations on a finite system we do not expect to observe an
abrupt phase transition, and there are also considerable computational
problems in reaching a steady state for temperatures too close to $T_{c}$,
and so we restricted our calculations to $T_{c}/T\leq 0.99$. The quantity
that we studied was the fraction $P(t)$ of sites at which the spin has not
changed by time $t$. For convenience we call $P(t)$ the relaxation function,
and the sites at which the spin has not changed as the unrelaxed sites. We
considered $P(t)$, rather than a spin auto-correlation function such as $%
C(t,t^{\prime })=<\sigma _{i}(t^{\prime })\sigma _{i}(t+t^{\prime })>$ with $%
\sigma _{i}\sigma _{j}=\delta (\sigma _{i},\sigma _{j})$,\ because the
latter contains a lot more random noise which will be strongly affected by
the value of $q.$ All the sites with no neighbors having the same spin were
omitted from the calculation of $P(t)$, since for these $E=0$ so that any
change of spin has $\Delta E\leq 0$ and the relaxation on these sites would
introduce into $P(t)$ a component that relaxed rapidly with time in a simple
exponential manner. This component corresponds to a fast $\beta $-relaxation
in a supercooled liquid, while the transitions that we consider resemble the
much slower ones of the $\alpha $-relaxation, in which we are interested
here. The question of the possible merging of the $\alpha $ and $\beta $
relaxations at high temperatures will be discussed elsewhere.

\section{Results of the calculations}

Extensive calculations were performed for $q=3,4,5,6,8$ on a lattices of 200
by 200 sites with periodic boundary conditions, a size that was adequate to
give very similar results for different sets of runs. For each value of $q$
and $T$, an array of random spins was first annealed until the system's
energy converged and then annealed further until the spins at 99\% of the
sites had changed. For high values of $T_{c}/T$, additional anneals were
performed as necessary until a steady state was obtained. After this, five
runs were performed on different initial states until $P(t)$ reached $0.01$,
and the average value of $P(t)$ was fitted to a stretched exponential
function, equation (1) \ This fit was excellent not only on a double log
plot, where many types of function appear as a straight line, but also on a
graph of $\ln [P(t)]$ as a function of $t$, except for occasional slight
deviations at very short or very long times. The results of the individual
runs, which contained much more random noise, were also fitted to stretched
exponential functions, and we checked that the average values of $\ln (\tau )
$ and of $\beta $ from the five runs were close to those obtained from
fitting the average value of $P(t)$. Accordingly, all the results presented
below relate to the values of the quantities from the average of five runs.
Incidentally, the scatter in the results increased with increasing $q$, as
is to be expected since the number of possible states in the system is $%
200^{2q}$. It also increased with increasing $1/T$, and was much greater for 
$\beta $ than for $\ln (\tau )$. For our calculations, it proved convenient
to choose $w_{0}=0.5/(q-1)$, so that the maximum total transition
probability at each step was the same, $0.5$, for all the systems.
Calculations were also performed for $q=2$ (the usual Ising ferromagnet),
but their results are somewhat different and so are not reported here. The
reason for this difference, as discussed in section 4, is that for $q=2$ any
spins differing from those at a given site are equal to each other, and this
affects the transitions rates for sites with $z=2$.

In figure 1 we present the values of $\ln (\tau )$ as functions of $1/T$ for
all the systems. In presenting the results, we choose $J=1/k_{B}$, so that $%
J/(k_{B}T)=1/T$. We only report here the results for $T_{c}/T\geq 0.5$,
since this contains the temperature range of main interest for the $\alpha $%
-relaxation. It is immediately apparent that the temperature \ dependence of 
$\ln (\tau )$ for all values of $q$ differs considerably from an Arrhenius
behavior, which would correspond to a straight line. While this temperature
dependence does not correspond to any of the standard functions used for
supercooled liquids \cite{Angell-rev}, it is qualitatively similar to that
found for such liquids. We found that the results all fit well on to the
master curve of $\ln (\tau )$ as a function of $T_{c}/T$ shown in figure 2
(apart from very close to $T_{c}$), and so we show all our other results as
a function of $T_{c}/T$ rather than of $1/T$. The values of $J/(k_{B}T_{c})$
are 1.005 for $q=3$, 1.10 for $q=4$, 1.174 for $q=5$, 1.238 for $q=6$, and
1.342 for $q=8$. In figure 3, we show the stretching exponent $\beta $ as a
function of $T_{c}/T$, and note at once that (at least for $T_{c}/T\leq 0.95$%
) the values of $\beta $ decrease steadily as $T_{c}/T$ increases (and so
the temperature decreases). 

As we discuss in the next section, the interpretation of these results is
connected with changes in the environments of the sites, and so we now
present the results of our calculations on them. We found two useful ways of
describing these environments. The first is in terms of the average value $%
<z>$ of the number $z$ of adjacent sites having the same spin as that of the
given site, which we refer to as the spin correlation. It is convenient to
divide $<z>$ by the coordination number, $z_{c}=4$, so that its maximum
value is unity, and we call the resulting quantity the reduced spin
correlation $sc$. The other quantity that we examined is the fraction of
sites inside clusters of the same spin (and not on their boundaries), so
that the spin on the site is the same as that on all its adjacent sites, and
we denote this quantity by $cl$. Since the sites in clusters all contribute
unity to $sc$, the difference $sc-cl$ is associated with sites on the edges
of clusters or outside them. The importance of the environment for the
relaxation rates is associated with its relationship to the probability for
a change of spin, as discussed in the next section.

The major properties that are of interest with regard to the environments of
the sites are how these change with temperature and with time. In order to
show the change in the environment as a function of time, we present in
figures 4 and 5 $sc$ and $cl$ as functions of the fraction $P$ of unrelaxed
sites\ for $T_{c}/T=0.9$ for all five values of $q$. As an example of how
this change in time varies with temperature, we present in figure 6 $cl$ as
a function of $P$ for $q=5$ and three different values of $T_{c}/T$, 0.7,
0.8 and 0.9, for which the values of the stretching exponent $\beta $ were
0.86, 0.79 and 0.72 respectively. In these figures, we chose $P$ to decrease
from 1 to 0 on the $x$-axis, as this is the direction of increasing time.\
Finally, in order to show the change in environment as a function of
temperature, we show in figures 7 and 8 the steady state values of $sc$ and $%
cl$ as functions of $T_{c}/T$.

\section{Discussion}

A comparison of figures 1-3 with figures 4-8 shows clearly that both
non-exponential relaxation and non-Arrhenius temperature dependence are
associated with the change in the average environment of the sites with
temperature and time, as we suggested in the Introduction. We now consider
in turn the temperature dependence and dependence on $q$ of the environments
of the sites, the relationship between the transition rates and the
environments, and the time dependence of the environments. After this, we
discuss the implications of our results for understanding the properties of
supercooled liquids.

The departure from an Arrhenius law of the temperature dependence of $\ln
(\tau )$ is associated with the increase as the temperature is lowered of
the the reduced spin correlation $sc$ and of the fraction of states inside
clusters $cl$. The reason for these increases is that the energy associated
with a site decreases as its spin correlation number $z$ increases, while
the number of possible arrangements of the spins associated with such a
state decreases, which leads to a decrease on the system's entropy. As the
temperature is lowered the role of entropy in the free energy becomes less
important, and this leads to an increase in $<z>$ and so in $sc$, as well as
to an increase in the fraction $cl$ of sites within clusters where their
energy is lowest. \ In order to obtain a better qualitative understanding of
this temperature dependence, let us suppose that the environments of the
different sites could all be chosen independently. If $C_{1}(z)$ denotes the
number of possible arrangements of spins round a given site, then the single
site partition function would be 
\begin{equation}
Z_{1}(T)=\sum_{z}C_{1}(z)\exp [zJ/(k_{B}T)],  \tag{4}
\end{equation}%
and the fraction of sites with a given value of $z$ would be%
\begin{equation}
n_{1}(z,T)=C_{1}(z)\exp [zJ/(k_{B}T)]/Z_{1}(T).  \tag{5}
\end{equation}%
If the sites with $z=0$ are included in the sum in equation (4), it can
readily be shown that for the square lattice $Z_{1}(T)=\{q-1+\exp
[J/(k_{B}T)]\}^{4}$. The fraction of states in a cluster in this
approximation is $1/Z_{1}(T)$, which increases as $T$ decreases and
decreases as $q$ increases. Since the arrangement of spins around adjacent
sites is obviously not independent, the entropy of the system cannot be
calculated just in terms of these single site quantities $C_{1}(z)$, but
equations (4)-(5) are useful for qualitative arguments such as the above. In
particular, they explain not only the increase in the fraction $cl$ of
particles in clusters as the temperature is lowered, as shown in figures 6
and 8, and the increase in the reduced spin correlation function $sc$ shown
in figure 7, but also why an increase in the number of possible values $q$
of the spin on a site leads to a reduction in the values of $cl$ and $sc$,
as found in figures 4, 5, 7 and 8.

We now turn to the relationship between the environment of the sites as
characterized by the functions $sc$ and $cl$ and the transition rates. For
sites inside a cluster, any transition reduces $z$ from $z_{c}$ to zero, and
so requires the maximum possible activation energy $E_{act}=z_{c}J=4J$, so
that $cl$ describes the fraction of sites at which a transition requires
this activation energy. For a site with $z=3$, there is one transition that
reduces $z$ from $3$ to $1$\ and so requires an activation energy of $2J$,
while any other transition requires an activation energy of $3J$. For sites
with $z\leq 1$ transitions are possible that do not require an increase in
energy, while for $z=2$ such transitions are only possible if the
neighboring sites contain two pairs of equal spins. Incidentally, this is
always the case for $q=2$, which is why the results for this value of $q$
differ somewhat from those for $q>2$ and so we decided not to report them
here. For the sites with values of $z$ that require thermal activation for
the transitions, an alternative relaxation mechanism is that the value of $z$
at that site decreases as a result in changes in the spin at adjacent sites
until unactivated transitions (or ones with a lower activation energy) are
possible. In order to find out which process is actually the dominant one,
we performed a number of calculations of the fraction of sites initially
having different values of $z$, the fraction at which the first transition
occurs for a given value of $z$, and the fraction at which the first
transition involves a given activation energy. In a typical set of results,
for $q=5$ and $T_{c}/T=0.9$, we found that 11\% of the sites have $z=0$,
24\% have $z=1$, 30\% have $z=2$, 24\% have $z=3$, and 11\% have $z=4$.
During the relaxation process we found that the neighborhoods of the
unrelaxed sites change, since at only 2\% of the sites did the first
transition occur with $z=4$, i.e. at sites within a cluster, while at 10\%
it occurred when $z=3$, at 34\% when $z=2$, at 38\% when $z=1$, and at 16\%
when $z=0$. A comparison of these results with the values of $z$ for the
whole system shows that the dominant mechanism for the relaxation of sites
with high spin coordination number $z$ is for the spins on some of the
adjacent sites to change before they relax, so as to reduce the activation
energy required for a transition. This is especially clear for sites
initially within a cluster (for which $z=4)$, since 9/11 of these sites had
lower values of $z$ when the first transition took place. As a result of
these changes of environment, 55\% of the first transitions did not require
any activation energy, 24\% required an activation energy of $J$, 14\% an
activation energy of $2J$, 5\% an activation energy of $3J$, and 2\% an
activation energy of $4J$. Such a behavior was found for all of the values
of $T_{c}/T$ that we studied (up to $T_{c}/T=0.99$). The time required for
this process is longest for sites within clusters, and increases with
increasing cluster size, while the average cluster size increases with
increasing $cl$, and so as the temperature decreases. This is one of the
main sources of the departure from an Arrhenius law temperature dependence
of the mean relaxation time.

With regard to the time dependence of the relaxation, those sites with a
lower initial value of $z$ will have the fastest relaxation rates and so
tend to relax first, while relaxation at the other sites will take longer as
it requires either a larger activation energy or a change in their value of $%
z$ as a result of transitions of their neighbors. As a result, $sc$ and $cl$
for the unrelaxed sites increases with time, as shown in figures 4-6, and
this leads to the observed departure from simple exponential relaxation. As
can be seen from figure 6, as $T_{c}/T$ increases the change in the
environment (as measured by $cl$) increases, and this correlates with the
decrease in $\beta $, which reinforces our conclusion that the
non-exponential time dependence is associated with changes in the
environment of the sites.

We now turn to some of the implications of our results for real supercooled
liquids. Our results strongly suggest that in real supercooled liquids the
temperature and time dependence of the relaxation are both associated with
changes in the average environments of the molecules, which explains why
they are often connected. This implies that a key factor in determining the
properties of supercooled liquids is the number of different environments
available for the molecules. Since this is smallest for covalently bonded
materials such as SiO$_{2}$, this explains why these are the strongest
liquids, with the smallest deviations from an Arrhenius law temperature
dependence of their relaxation times and from a simple exponential time
dependence of their relaxations. On the other hand, there are numerous
possible molecular environments for van-der-Waals bonded liquids, which
explains why these tend to be the most fragile and to have the lowest
stretching exponents for their relaxation. H-bonded liquids have an
intermediate number of possible molecular environments, because of the
restrictions imposed by the hydrogen bonds, and so their fragility and
stretching exponents tend to be intermediate between those of covalently
bonded and van-der-Waals bonded liquids. In addition, since the glass
transition involves a freezing of the molecular environments, the decrease
in entropy when this happens will increase with the number of possible
environments, and so with the fragility, as is found experimentally. A more
detailed analysis of these effects is beyond the scope of this paper. Our
results also explain qualitatively the connection between fragility and the
degree of non-exponential relaxation observed in many supercooled liquids %
\cite{Ngai-rev}. The effects of changes in the environment on the lifetimes
of dynamic heterogeneities found in supercooled liquids is most easily
discussed if one associates these heterogeneities with molecules inside
clusters, which are often assumed to be solid-like. These clusters are the
states of lowest local energy, so that an increasing fraction of molecules
occupy them as the temperature is lowered. In principle, just as for the
Potts model, such molecules can relax either by a particle in the middle of
a cluster acquiring enough activation energy to change its state or
particles on the edge of a cluster relaxing in turn so that the clusters of
unrelaxed particles gradually disintegrate. These two processes resemble,
respectively, the formation and motion of interstitial atoms in a crystal
and the dissociation of subcritical nuclei in a liquid. If the latter
process is more probable, as one would expect at temperatures above the
glass transition temperature just as it was found to be for the Potts model
at temperatures above $T_{c}$, then the lifetimes of the dynamic
heterogeneities will be of the same order of magnitude as the $\alpha $-
relaxation time, as is found experimentally in many liquids \cite{Russell}%
\cite{Sillescu}.

\section{Conclusions}

One major advantage of the Potts model studied in this paper over many other
models is that it does not require the imposition of any external
assumptions, such as the existence of liquid-like and solid-like regions or
of dynamic restrictions, although these may well be present in the model.
Instead, all the results are derived directly from the system's Hamiltonian.
The results of our calculations on the two-dimensional ferromagnetic Potts
model show that both the temperature dependence and the time dependence of
the relaxation are associated with changes in the arrangements of spins
around the different sites, i.e. the environments of the sites. Such a
feature can explain why covalently bonded liquids, in which there is only a
small number of possible molecular environments, are strong, while
van-der-Waals bonded liquids, in which this number is large, tend to be the
most fragile. It also explains the connection between stretched exponential
relaxation and a non-Arrhenius temperature dependence of the relaxation time
in supercooled liquids. In addition, the relaxation in the Potts model of
spins within clusters of sites by the disintegration of the clusters
suggests that a similar mechanism is relevant to the relaxation of molecules
inside dynamic heterogeneities, and this would explain why the lifetimes of
such heterogeneities is usually found to be of the same order of magnitude
as the $\alpha -$relaxation time.

Figure captions.

Figure 1.(Color on line) The relaxation time $\tau $ of the stretched
exponential relaxation of $P(t)$ as a function of $1/T$ for $q=3$ (circles,
red), $q=4$ (stars, violet), $q=5$ (squares, black), $q=6$ (hexagons,
magenta) and $q=8$ (triangles, blue). The dotted lines are just to guide the
eye.

Figure 2. (Color on line) The relaxation time $\tau $ of the stretched
exponential relaxation of $P(t)$ as a function of $T_{c}/T$ for $q=3$
(circles, red), $q=4$ (stars, violet), $q=5$ (squares, black), $q=6$
(hexagons, magenta) and $q=8$ (triangles, blue). The dotted lines are just
to guide the eye.

Figure 3. (Color on line) The stretching exponent $\beta $ of the stretched
exponential relaxation of $P(t)$ as a function of $T_{c}/T$ for $q=3$
(circles, red), $q=4$ (stars, violet), $q=5$ (squares, black), $q=6$
(hexagons, magenta) and $q=8$ (triangles, blue). The dotted lines are just
to guide the eye.

Figure 4. (Color on line) The reduced correlation function $sc(P)$ as a
function of $P$ for $T_{c}/T=0.9$ for $q=3$ (red), $q=4$ (violet), $q=5$
(black), $q=6$ (magenta) and $q=8$ (blue).. The values of $q$ are shown on
the curves.

Figure 5. (Color on line) The fraction of sites in clusters $cl(P)$ as a
function of $P$ for $T_{c}/T=0.9$ for $q=3$ (red), $q=4$ (violet), $q=5$
(black), $q=6$ (magenta) and $q=8$ (blue).. The values of $q$ are shown on
the curves.

Figure 6. The fraction of sites in clusters $cl(P)$ for $q=5$ as a function
of $P$ for $T_{c}/T=0.7$.(dashes, blue), $T_{c}/T=0.8$.(dot-dash, red) and $%
T_{c}/T=0.9$.(solid, black). The values of $T_{c}/T$ are marked on the
curves.

Figure 7. (Color on line) The steady state reduced correlation function $%
sc_{0}$ as a function of $T_{c}/T$ for $q=3$ (circles, red), the top curve, $%
q=4$ (stars, violet), $q=5$ (squares, black), $q=6$ (hexagons, magenta) and $%
q=8$ (triangles, blue), the bottom curve. The dotted lines are just to guide
the eye.

Figure 8. (Color on line) The\ steady state fraction of sites in clusters $%
cl_{0}$ as a function of $T_{c}/T$ for $q=3$ (circles, red), the top curve, $%
q=4$ (stars, violet), $q=5$ (squares, black), $q=6$ (hexagons, magenta) and $%
q=8$ (triangles, blue), the bottom curve. The dotted lines are just to guide
the eye.


\begin{thebibliography}{99}
\bibitem{Angell-rev} C. A. Angell, K. L. Ngai, G. B. McKenna, P. F.
McMillan, S. W. Martin, J. Appl. Phys. \textbf{88},\textbf{\ }3114 (2000)

\bibitem{Angell-fragile} C. A. Angell, J. Phys. Chem. Solids \textbf{49},
863 (1988)

\bibitem{Ngai-rev} K.L. Ngai, J. Non-Cryst. Sol. \textbf{275}, 7 (2000)

\bibitem{Wu-rev} F. Y. Wu, Rev. Mod. Phys. \textbf{54 }235 (1982)

\bibitem{FA} G.H. Fredrickson and H.C. Andersen, J. Chem.Phys. \textbf{83}
5822 (1985)

\bibitem{Benkhof} S Benkhof, A Kudlik, T Blochowicz and E Rossler, J. Phys:
Condens. Matter \textbf{10} 8155 (1998)

\bibitem{Bermejo} F J Bermejo, M Jimenez-Ruiz, A Criado, G J Cuello, C
Cabrillo, F R Trouw, R Fernandez-Perea, H Lowen and H E Fischer, J. Phys:
Condens. Matter \textbf{12} A391 (2000)

\bibitem{Brang} C Brangian, W Kob and K Binder, J. Phys. A: Math. Gen. 
\textbf{36} 10847 (2003)

\bibitem{Velytsky} A. Velytsky, B. A. Berga and U. M. Hellera, Nuclear
Physics B (Proc. Suppl.) \textbf{119} 861 (2003)

\bibitem{Ngai-coupling} K L Ngai, Comments Solid State Phys. \textbf{9} 121
(1979)

\bibitem{Bendler} J. T. Bendler, J. J. Fontanella and M. F. Shlesinger, J.
Chem. Phys. \textbf{118} 6713 (2003)

\bibitem{Tanaka} H Tanaka, J. Non-Cryst. Sol. \textbf{351 }3385\textbf{\ }%
(2005)

\bibitem{Garrahan} J. P. Garrahan and D. Chandler, Phys. Rev. Lett. \textbf{%
89}.035704 (2002); Berthier L and Garrahan J P, J. Chem. Phys. \textbf{119}
4367 (2003).

\bibitem{Gotcheva} V. Gotcheva, Y. Wang, A. T. J. Wang and S. Teitel, Phys.
Rev. \ B \textbf{72}, 064505 (2005).

\bibitem{Baxter} R. J. Baxter, J. Phys. C: Solid State Phys. \textbf{6} L445
(1973).

\bibitem{Russell} E V Russell and N E Israeloff, Nature \textbf{408} 695
(2000)

\bibitem{Sillescu} H Sillescu, R Bohmer, G. Diezemann and G. Hinze, J.
Non-Cryst. Sol. \textbf{307-310} 16 (2002)
\end{thebibliography}
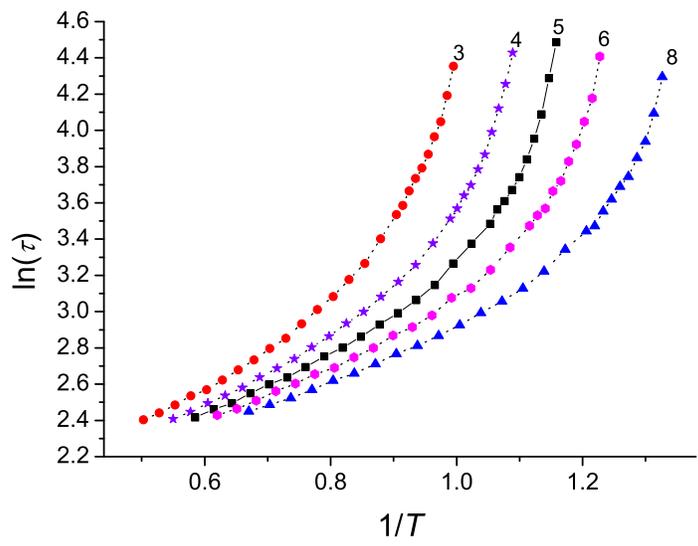
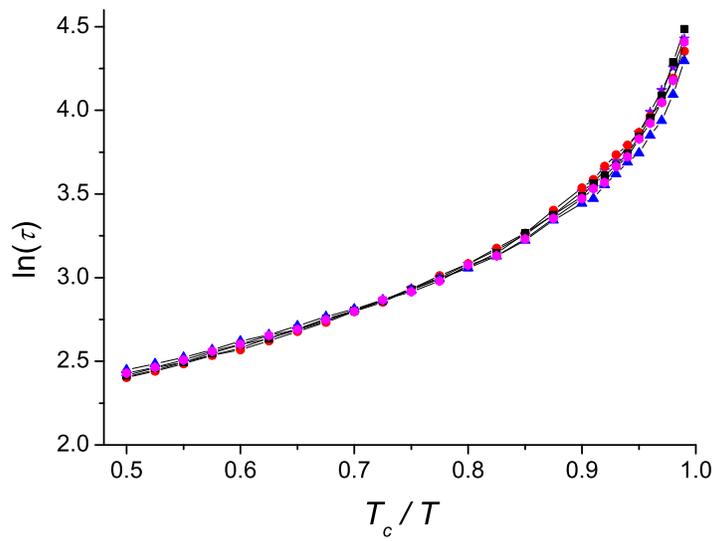
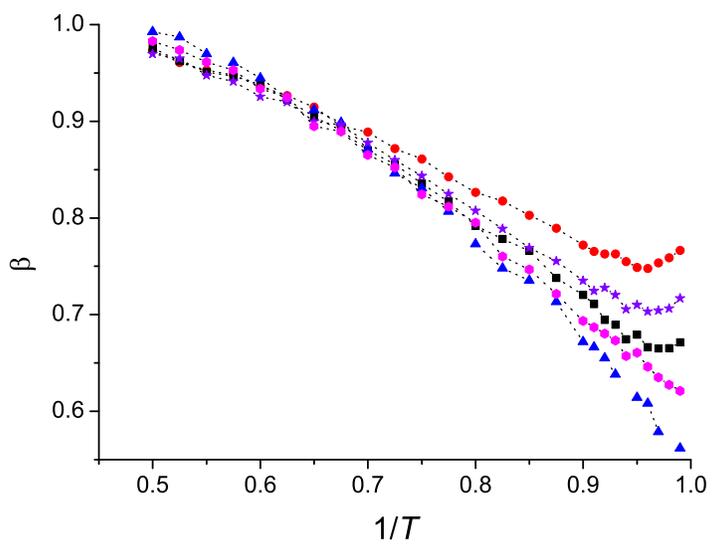
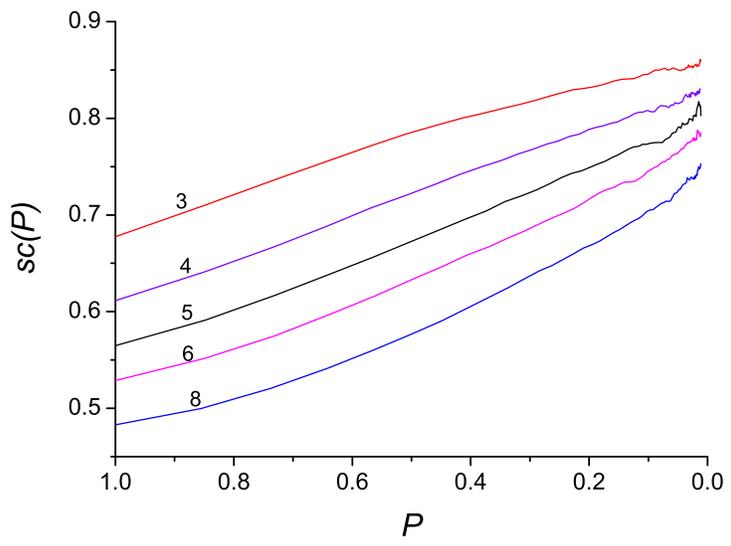
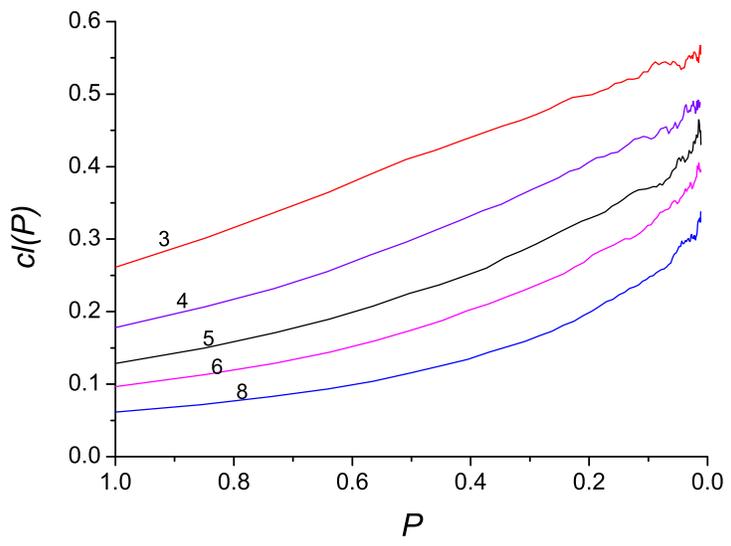
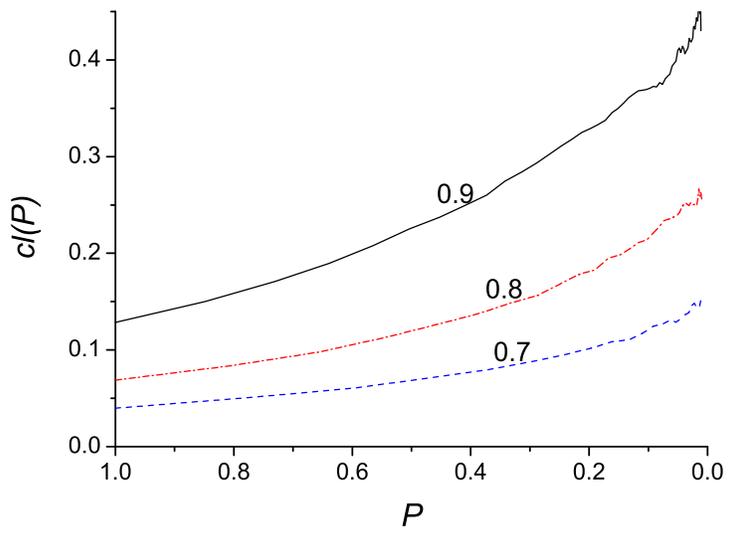
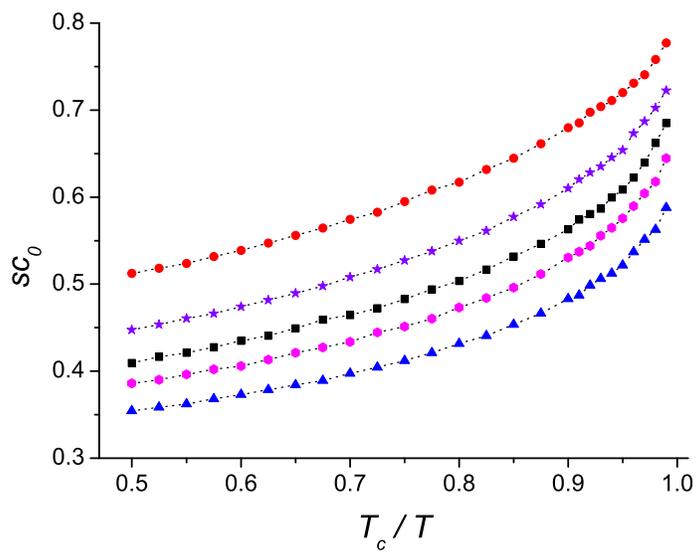
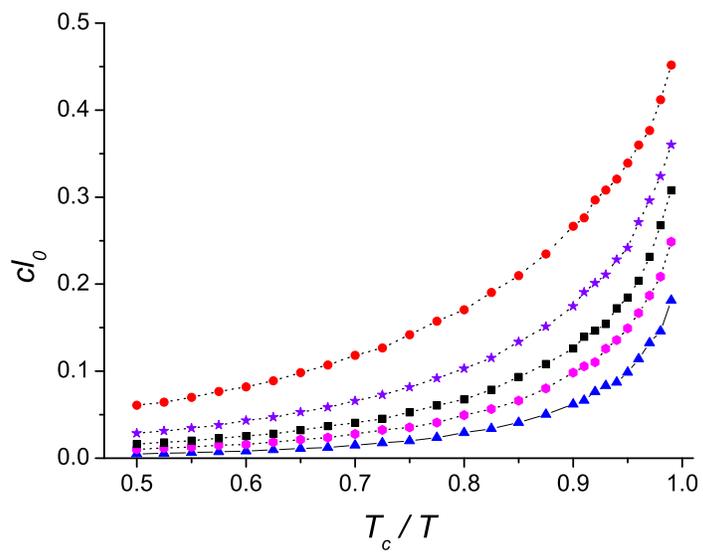
\end{document}